\documentclass[aps, pra, a4paper, showpacs, twocolumn, english, 10pt]{revtex4-2}
\usepackage{bbm, amsthm, bm,textcomp, nicefrac,geometry,ragged2e}
\usepackage[dvipsnames]{xcolor}
\usepackage{float}
\usepackage[bbgreekl]{mathbbol}
\usepackage{graphicx,epstopdf,color,verbatim,enumitem,ulem}
\usepackage{pbox,array}
\usepackage{mathrsfs}
\usepackage{amssymb}
\usepackage{textgreek}
\usepackage{mathtools}
\usepackage{stackrel}
\usepackage[thinlines]{easytable}
\usepackage{amsmath}
\usepackage[utf8]{inputenc}
\makeatletter
\usepackage{soul}
\usepackage[caption=false]{subfig}
\usepackage{hyperref}
\hypersetup{
    colorlinks = true,
    linkcolor = Red,
    citecolor = Red,
    filecolor = Black,
    urlcolor = Black,
}
\usepackage[T1]{fontenc}
\usepackage[caption=false]{subfig}
\usepackage{babel}
\usepackage[linesnumbered,ruled]{algorithm2e}

\addto\captionsenglish{}

\definecolor{brickred}{rgb}{0.8, 0.0, 0.0}

\begin{document}

\title{Enhancing quantum processor capabilities during idle times}

\author{Wolfgang D\"ur$^1$}

\affiliation{$^1$Universit\"at Innsbruck, Institut f\"ur Theoretische Physik, Technikerstra{\ss}e 21a, 6020 Innsbruck, Austria}

\date{\today}

\begin{abstract}
We advocate an alternative paradigm to operate quantum computers that utilizes multipartite entanglement generated in dedicated auxiliary systems during idle times. This stored entanglement enhances the future capabilities of the quantum processor, as it can be flexibly used to assist and speed-up computations when needed. We identify classes of multipartite entangled resource states whose computational power are related to their entanglement features, and in turn to the complexity to generate them. During idle times, one can thus continuously work towards generating more and more powerful auxiliary multipartite entangled states. Idle times include both times prior to the start of a computation, but also any step during the execution of an algorithm where parts of the processor are not actively involved. To illustrate our approach, we consider architectures with limited connectivity, e.g. corresponding to a 1D geometry. We show that $d$-dimensional cluster states allow one to flexibly perform multiple long-distance two-qubit gates in parallel, where both the complexity to generate them, as well as the number of achievable gates increases with $d$.

\end{abstract}

\maketitle

\section{Introduction}
Entanglement is a valuable resource that enables one to perform tasks that are impossible otherwise, ranging from security application and quantum cryptography to enhanced quantum sensing. In the context of quantum computation, entanglement is long believed as the key for quantum speedup. This is particularly highlighted in the paradigm of measurement-based quantum computation \cite{Raussendorf2001,Raussendorf2003,Briegel2009}, where a highly entangled resource state, a so-called 2D cluster state \cite{Briegel2001}, is processed solely by single qubit measurements. In recent years, it was realized that entanglement, together with measurements, can also support circuit-based quantum computation. The advantage of mid-circuit measurements and measurement-assisted methods for state preparation and computation has been analyzed in multiple works, see e.g. \cite{Hillmich2021,RN12,Baeumer2024a,PhysRevApplied.23.014057,Plathanam2025,Baeumer2024,RN9,RN13,RN14,RN15,RN16,RN17,RN18,Beverland2022,yang2023,Piroli2024,PRXQuantum_6_010306,choe2024,Devulapalli2024,Yu2025Dicke,Kaldenbach_2025,Devulapalli2025a}.

What these methods have on common is that they utilize measurements in addition to quantum gates, as part of a certain preparation task or algorithm. In particular, entanglement is generated during this process on demand.

Here we go one step further and point out that possibility to support and accelerate quantum computations by utilizing entanglement in fact leads to a new paradigm how to design and operate quantum computers. We show that with an appropriate ancilla-assisted architecture \cite{Eldredge2020,rierasabat2024,Dur2025}, quantum processor idle times can be used to enhance the future capabilities of the quantum processor. The available quantum systems are thereby separated into data- and auxiliary systems, where during processor idle times, i.e. when no quantum computation is performed, or when parts of the processor are not actively used in an ongoing computation, (mutipartite) entangled states are generated or upgraded among the auxiliary systems. These states can later be flexibly used to assist and accelerate computations, thereby making the quantum processor more powerful.
This is particularly useful whenever there are some geometric constraints in the set-up, for instance w.r.t. available connections (such as only nearest neighbor couplings), or in terms of gate parallelization. We show how these limitations, which otherwise lead to computational overheads in terms of run-time, can be overcome with our entanglement-assisted approach.

\begin{figure}[ht]
    \centering
    \includegraphics[width=\columnwidth]{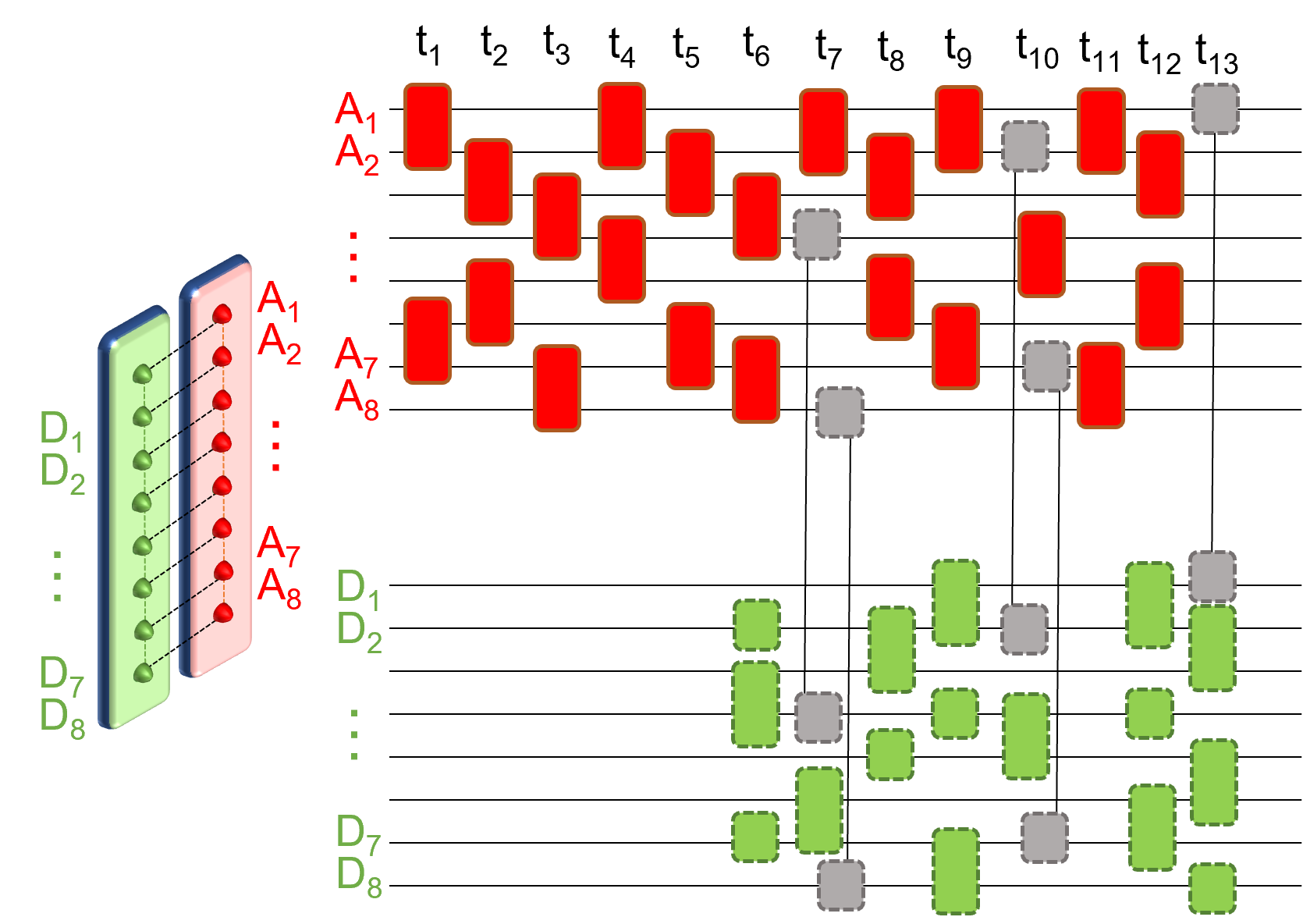}
    \caption{\label{Fig_idle} Illustration of the usage of idle times to enhance the performance of a quantum computer. Qubits are separated into data qubits $D$ (green) and auxiliary qubits $A$ (red), where dashed lines indicate available interactions. The quantum computation takes place on the data qubits in time steps $t_6$ to $t_{13}$, while (multipartite) entangled states are generated among the auxiliary qubits whenever possible. This entanglement is utilized to perform teleportation-based gates between distant data qubits on demand, here in time step $t_7$ on $D_4D_8$ and in time step $t_{10}$ on $D_2D_7$. Such long-distance gates would otherwise require multiple time steps due to nearest-neighbor interaction limitations. Whenever part of system $A$ is not involved in the actual quantum computation, entanglement is generated and enhanced. These idle times involve times prior to the start of the actual computation ($t_1 \ldots t_5$), but also time steps during the computation - even when parts of the auxiliary systems assist the computation. The more idle time is available, the more powerful the possible assistance due to auxiliary entanglement.
   }
\end{figure}

In particular, we consider multipartite entangled states as resource, which can be flexibly manipulated by local measurements, and used to assist quantum computations. Importantly, these states can be generated beforehand, and independent of the particular algorithm or task that should be performed later. This is different from the on-demand generation of specific entangled states to enhance particular algorithms or tasks \cite{Eldredge2020, Beverland2022, yang2023, Devulapalli2024, Devulapalli2025a, Plathanam2025, Kaldenbach_2025}. What is more, we identify classes of resource states with different computational power. We show that $d$-dimensional cluster states (dD-CSs) allow one to flexibly assist computations, where the obtained enhancement is improved for larger $d$. In turn we discuss how to efficiently generate and upgrade such dD-CSs.

This leads to an alternative paradigm to perform quantum computations: Before the quantum processing starts, or at any stage where parts of the quantum processor are not participating in the execution of an algorithm (e.g. since in a given step only a subset of auxiliary qubits is utilized for the computation), one works towards generating more and more powerful auxiliary entangled states. This can be done either by generating and storing new states, or by upgrading the type of entangled states that are stored, e.g. by working towards generating dD-CSs with increasing dimension $d$. This highlights the active role of a quantum memory in a quantum computation, which vastly exceeds the one of its classical counterpart in classical computation. A quantum memory can be used not only to passively store quantum data, but also to actively support quantum information processing.

The main contributions of this work are as follows:
\begin{itemize}
\item We analyze a quantum processor architecture with dedicated data- and auxiliary systems, and highlight the role of auxiliary (multipartite) entanglement as valuable resource to flexibly support and accelerate quantum computations in set-ups with (geometric) constraints.
\item We identify $d$-dimensional cluster states as useful resources, and show that they can e.g. be utilized to flexibly perform multiple two qubit gates in parallel, where the number of gates increases with the dimension $d$ as  ${\cal O}(N^{(d-1)/d})$.
\item We show how to efficiently generate and upgrade $d$-dimensional cluster states utilizing mid-circuit measurements in set-ups with limited connectivity, including 1D and 2D interaction geometries.
\item We argue that the computational power of quantum processors can be gradually increased if more idle time is available, either before a computation starts, or during a computation.
\end{itemize}

We illustrate our approach by considering settings with limited interaction geometries, where e.g. quantum interactions are restricted to a short range, such as nearest-neighbor quantum interactions in a one-dimensional or two-dimensional spatial arrangement of the qubits. Such restrictions occur naturally in several physical realizations, e.g. quantum computers based on superconducting qubits \cite{RN26,RN27,RN28}, but also in some setups based on trapped atoms \cite{RN29,RN30,RN31,RN32} or ions \cite{RN19,RN20,RN21,RN22,RN23}. In such settings, long-distance gates may be realized only with significant overhead, as multiple short-distance gates are required to realize a long-distance one. We also consider systems with bus-based interactions, where there is no geometric restriction on available gates, but on their parallelization - i.e. where gates can only be performed sequentially \cite{Cirac&Zoller1995, fazio2025}.
Here we illustrate the computational advantage offered by auxiliary entangled states by their capability to realize a certain number $m$ of such long-ranged gates, where a larger number $m$ of gates that can be done in parallel corresponds to a larger computational power.
We illustrate of our approach with help of dD-CSs, where the number $m_d$ of long-ranged gates that can be implemented increases with $d$.  We emphasize, however that this is meant as an example, and we anticipate that other state families and supporting tasks enabled by utilizing multipartite entanglement can be found.

The paper is structured as follows. In Sec. \ref{Sec_Setup}, we describe the set-up and the quantum computer architecture we consider, where the key element is the separation of the system into a part that is responsible to store and process quantum information, and an auxiliary system to generate and store entanglement. The latter allows one to perform certain tasks and support the quantum computation on demand. We also introduce an underlying 1D interaction geometry as an illustrative example to showcase the advantage of utilizing multipartite entanglement to overcome intrinsic limitations, which otherwise lead to overheads.  In Sec. \ref{Sec_Background} we review graph states and their manipulation, and how entanglement can be utilized to perform certain tasks, e.g. two- and multi-qubit quantum gates between remote qubits. Sec. \ref{Sec_Bellstateextraction} is concerned with the flexible extraction of multiple Bell states from cluster states of different dimension, where we show that cluster states corresponding to lattices of higher spatial dimension allow one to generate more Bell pairs. In Sec. \ref{Sec_advantage} we discuss the advantage different kinds of multipartite entangled states offer a quantum processor for different tasks, including performing multiple two-qubit gates between remote qubits, as well as realizing general permutations, or Clifford circuits. We then turn to the measurement-assisted generation of $d$-dimensional cluster states in a 1D interaction geometry in Sec. \ref{Sec_Stategeneration1D}.
We also show how to gradually increase the computational power of stored resource states. We discuss alternative resource states and the role of noise and imperfections in Sec. \ref{Sec_Summary}, where we also summarize and conclude.

\section{Setup and scheme}\label{Sec_Setup}
\begin{figure}[ht]
    \centering
    \includegraphics[width=\columnwidth]{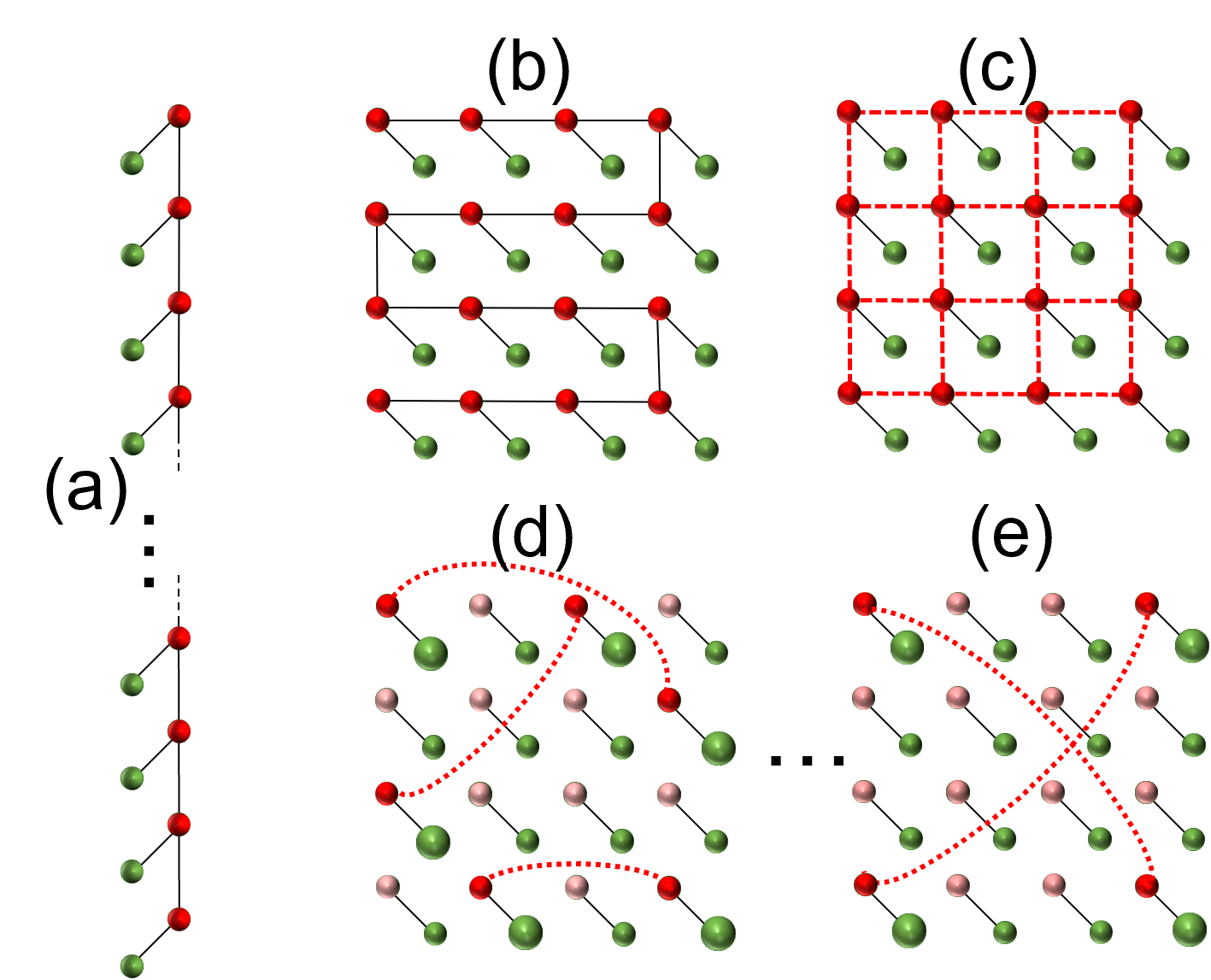}
    \caption{\label{Fig_2DClustertoBell} (a) 1D interaction geometry between auxiliary systems (red), where a data qubit is associated with each auxiliary qubit. (b) The qubits are virtually arranged in a snake-like way to resemble a 2D lattice. (c) By using a sequence of nearest-neighbor gates w.r.t. 1D geometry, one can generate a multipartite entangled state corresponding to a 2D lattice, i.e. a 2D cluster state. Notice that the obtainable entanglement topologies are independent from the interaction geometry. The 2D cluster state acts as a (universal) resource to assist quantum computations: it can be flexibly transformed by means of single qubit measurements into different patterns of multiple entangled Bell pairs, see (d) and (e) for two possible examples. Each of these Bell pairs allows one to perform a long-distance, teleportation-based two-qubit gate on the associated data qubits, and hence multiple such gates can be performed in a single time step on data qubits utilizing an auxiliary 2D cluster state.
       }
\end{figure}

We consider a quantum computational architecture with some given (restricted) interaction geometry, where the role of quantum systems is split into data systems (set $D$, green) and auxiliary systems (set $A$, red and blue) \cite{Dur2025,Plathanam2025}. Data systems are used to store and process quantum information, while auxiliary systems are used to generate and store auxiliary entanglement, which in turn is utilized to assist the computation, e.g. to perform teleportation-based gates between data qubits. For simplicity, we consider two-level quantum systems (qubits) and assume an underlying 1D interaction geometry, where auxiliary qubits are only coupled to their nearest neighbor, there is no direct coupling between data qubits, and each data qubit is coupled to at least one auxiliary qubit (see Fig. \ref{Fig_2DClustertoBell}a).
We will also consider other underlying interaction geometries later, including direct connections between data qubits in set $D$, a 2D interaction geometry in set $A$, or additional connections between the two sets.

Within this setting, single-qubit gates are performed directly on data qubits, while gates between data qubits are realized utilizing entangled states of the auxiliary systems via gate teleportation \cite{Eldredge2020,Hillmich2021,Baeumer2024a,Plathanam2025,Dur2025} \footnote{Notice that in a setting where also data qubits are connected and can interact according to some interaction geometry, the corresponding two-qubit gates between data qubits may be implemented directly.}. These entangled states can be generated on demand between entangling qubits \cite{Baeumer2024a,Plathanam2025}, or alternatively some (universal) multipartite entangled state can be generated, and utilized in a flexible way \cite{Dur2025}. Multipartite entanglement between the auxiliary systems can then be transformed by means of measurements to different forms, e.g. multiple Bell states shared between several pairs of qubits \cite{Freund2024,Dur2025}, see Fig. \ref{Fig_2DClustertoBell}. The desired target distribution of Bell states can thereby be flexibly chosen by properly adjusting the measurement basis. Each Bell state allows one to implement a two-qubit gate between the corresponding data qubits, while other types of graph states allow one to implement whole Clifford circuits on subsystems \cite{rierasabat2024}. The number $m$ of parallel Bell states --and hence parallel two-qubit gates-- that can be obtained in this way depends on the type of multipartite entangled state stored in the auxiliary system. Specifically, we will consider $d$-dimensional cluster states (dD-CSs) corresponding to $d$-dimensional lattices with $d=1,2,3,4, \ldots$ in the following. Our approach is however not limited to these state families. We find that the number parallel Bell states one can obtain increases with $d$, and is given by ${\cal O}(N^{(d-1)/d})$. Remarkable, generating multiple Bell states from a dD-CS and using them to implement multiple gates among data qubits can be done in a single step.

We also show how to efficiently generate dD-CSs using the underlying 1D interaction geometry among the auxiliary qubits, where generating cluster states with higher dimension $d$ requires more steps. The required resource states in the auxiliary system can be generated using multiple rounds of entangling gates. We demonstrate that this can be accelerated by allowing for additional qubits to store quantum states (set $S$, blue), and allow for mid-circuit measurements. Remarkably, these auxiliary states can be generated beforehand, i.e. before it is known which two-qubit gates should be performed. What is more, one can gradually increase the computational power of the resource states, by first preparing a 1D-CS, then work towards preparing a 2D-CS if more idle time is available, and continue in this way by preparing dD-CSs with increasing $d$. In this sense, the stored auxiliary states are a valuable resource to accelerate quantum computations, and their inherent power can be increased even while the quantum processor (or parts of it) is idle.

While we describe the method with bare qubits, we stress that the approach is also applicable in fault-tolerant quantum computation scenarios where logical systems and encoded quantum information is used, both for data qubits but also for auxiliary systems. In this case encoded entangled states are generated and utilized to (fault-tolerantly) process encoded quantum information.

\section{Background}\label{Sec_Background}

\subsection{Graph states and d-dimensional cluster states}
Throughout this paper, we consider a specific class of multipartite entangled states, so-called graph states \cite{RN7,heinEntanglementGraphstates}. Consider an arbitrary graph $G$, i.e. a set of vertices $V$ and edges $E$. The associated graph state of $n$ qubits is defines as
\begin{equation}
|G\rangle = \prod_{(i,j)\in E}CZ_{ij}|+\rangle^{\otimes n}.
\end{equation}
Therein, CZ$_{ij}$ is a controlled-Z gate (“CZ gate” for short) acting on auxiliary qubits $i$ and $j$, and $E$ is the edge set of the corresponding graph $G$, where the edge between qubits $i$ and $j$ is denoted as $(i,j)$. The CZ gate is defined by CZ=diag$(1,1,1-1)$, and we denote $|+\rangle=1/\sqrt{2}(|0\rangle + |1\rangle)$. In other words, the graph state is obtained by applying a product of CZ gates, one for each edge of the graph $G$, to the initial state $|+\rangle^{\otimes n}$.

Here we consider in particular graph states corresponding to rectangular lattices of dimension $d$, i.e. dD-CSs. In 1D, this corresponds to a 1D string with edges $(j,j+1)$, while in 2D we have a graph corresponding to a square lattice, and in 3D a cube structure.

\subsubsection{Graph state manipulation by means of single-qubit measurments}
Consider an arbitrary graph state corresponding to a graph $G$. If a qubit of the graph state is measured in the Pauli $X$, $Y$ or $Z$ basis, the resulting quantum state is (up to a local unitary operation) again a graph state, with a different graph \cite{heinEntanglementGraphstates}. Specifically, the action of a Pauli $X$, $Y$ or $Z$ measurement of a qubit located at a vertex $a$ of the graph $G$ is described, up to local unitary correction operations, by the following graphical rules:
	(i) a $Z$-measurement erases all edges connected to $a$;
	(ii) a $Y$-measurement first inverts the neighborhood graph of $a$ (the neighbors of vertex $a$ are all vertices connected to $a$ by an edge; the neighborhood graph of $a$ is the subgraph between the neighbors of $a$; inverting the neighborhood graph of $a$ means that any edge between two neighbors of $a$ is replaced by a non-edge, and vice versa) followed by a deletion of all edges connected to $a$;
	(iii) an $X$-measurement first inverts the neighborhood graph of an (arbitrary) neighbor $b$ of $a$, followed by an inversion of the neighborhood graph of $a$, followed by a deletion of all edges connected to $a$, followed by another inversion of the neighborhood graph of $b$.
The above graphical rules can be used to show how Bell states can be prepared between pairs of auxiliary qubits by performing single-qubit measurements on a dD-CS. The auxiliary qubits may be far away from each other, and thus the Bell states correspond to long-range entanglement across the lattice of auxiliary qubits.

\subsection{Remote two-qubit gates using Bell pairs}\label{Telgate}
Consider a Bell-type state $|B\rangle = CZ|+\rangle|+\rangle = 1/\sqrt{2}(|0\rangle|+\rangle + |1\rangle|-\rangle)$ shared among two auxiliary qubits. The state can be used to realize e.g. a two-body CZ gate on the associated data qubits by performing a teleportation-like protocol. To this end, the two data qubits are coupled to the state $|B\rangle$ via respective CZ gates. Thereafter, the two auxiliary qubits of the Bell pair $|B\rangle$ are measured in the $X$ basis. Depending on the measurement outcomes $(-1)^{m_k}$ with $m_1,m_2 \in \{0,1\}$, local $Z$-correction operations $Z^{m_2}\otimes Z^{m_1}$ on the data qubits are performed. This results in the application of a CZ gate to the two data qubits. This is similar -up to a local basis change- to the remote implementation of gates described in \cite{rierasabat2024,Baeumer2024a,Plathanam2025}.

It is easy to adapt the above procedure to realize, for example, a CNOT gate acting on the data qubits, since CNOT is equal to CZ up to the application of a single-qubit unitary gates (namely Hadamard gates performed on the target qubit before and after the CZ gate). Notice that the coupling between the auxiliary qubits and the data qubits can already be done before the 2D cluster state is processed to prepare the Bell state, and hence does not require an additional time step.

In light of the above, a large number $N$ quantum operations, which may be long-range gates, can be realized in parallel by processing each of the respective Bell states (and the associated data qubits) in the above-described manner, as depicted in Fig. \ref{Fig_2DClustertoBell}. 

\subsection{Multi-qubit gates, multiple CZ gates and Clifford circuits}\label{MultiTelgate}
As described in \cite{rierasabat2024} (see also \cite{yang2023}), one can utilize multipartite entangled states to implement different kinds of gates or even whole circuits. In particular, GHZ states allow one to perform gates of the form $\exp(-i\alpha Z^{\otimes m})$ with arbitrary $\alpha$, and a graph state can be used to simultaneously implement CZ gates between all pairs of qubits that are connected by an edge in the corresponding graph. Finally, by using stabilizer states of size $2m$ as resource states, i.e. with two qubits per node, one can realize whole Clifford circuits on $m$ qubits in a single time step. These resource states are local unitary equivalent to graph states. In all cases, the entangled resource state is coupled to the data qubits, and appropriate measurements implement the desired unitary operation or circuit on data qubits in a single step.

Gate-based methods to generate Clifford circuits are e.g. discussed in \cite{Bravyi2022}, where however the availability of long-ranged gates is assumed.

\section{Extraction of multiple Bell states from $d$-dimensional cluster states}\label{Sec_Bellstateextraction}
Here we discuss how a $d$-dimensional cluster state corresponding to a $d$-dimensional rectangular lattice can be transformed in a flexible way to multiple Bell states shared between $m$ pairs of qubits. Generally, a Bell pair can be generated between two qubits by identifying a path between them. Then, all qubits among the path (except the end points) are measured in $Y$, and all neighboring qubits among the path (and of the end points) are measured in $Z$ to isolate the resulting Bell state from the remaining structure. In 1D and 2D, this strategy allows one in general only to obtain a single Bell pair, since one cuts out a 1D chain which hinders further connections (crossing of pathes). For $d\geq 3$, the crossings can be avoided by choosing for the next Bell state a path that uses the third (or some other) dimension. Note also that for a fixed system size $N$, the average distance between any two points in a regular $d$-dimensional lattice is given by $l_d=\tfrac{d}{2}N^{1/d}$, i.e. decreasing significantly with $d$. Also all neighbors of qubits among the path and the endpoints need to be measured, such that the generation of a Bell state between two arbitrary qubits involves on average $2d(l_d+1) \approx d^2 N^{1/d}$ qubits, where the reduced number of neighbours at $d-1$ turning points are not taken into account. The remaining structure is still largely unaltered, since only a quasi-1D line is cut out from a $d$-dimensional lattice. Hence additional Bell states can be generated by using the remaining graph state. In total, one can transform the initial $d$-dimensional $N$-qubit cluster state to approximately
\begin{equation}
\label{m_d}
m_d=(N/(d^2N^{1/d})={\cal O}(N^{(d-1)/d})
\end{equation}
parallel Bell states, where we estimate the number of obtainable Bell pairs by the size of the system divided by the average number of resources (i.e. measured qubits) per Bell pair. Notice that one can be flexible choose which pairs of qubits should form a Bell pair. Almost all target configurations are achievable \footnote{Notice that some configurations might not be obtainable, e.g. if end-points of different Bell states are too close to each other, or are located among another path. In the latter case, choosing an alternative path could circumvent the problem though.}. 
We remark that in 2D, an alternative strategy, the zipper scheme based on $X$ measurements among staircase pathes, was proposed in \cite{Freund2024}, also leading to ${\cal O}(\sqrt{N})$ parallel Bell states that can be generated (see also \cite{Dur2025}).

We also remark that from a 2D cluster states, also other types of target states including arbitrary graph states of size $m$ on a subset of qubits can be generated, see \cite{Freund2025}. These methods can be extended to dD-CSs with $d\geq 3$, where a more efficient generation of target graph states as compared to 2D-CSs can be expected.

\section{Enhanced computational power of quantum processor}\label{Sec_advantage}
In a quantum processor with a given interaction geometry, there is a certain overhead associated with different tasks. These overheads depend on the particular set-up, e.g. the available interaction geometry \cite{cowtan2019,childs2019,Bapat2021,Bapat2023,Yuan_2025}, how gates and measurements are realized, and whether gates can be done in parallel. Entanglement may allow for a more efficient realization of certain tasks and reduce these overheads, and in this sense acts as a resource \cite{Eldredge2020, yang2023, Beverland2022, Baeumer2024a, Devulapalli2024, Devulapalli2025a, Plathanam2025,Kaldenbach_2025, Dur2025}. Any such speedup is relevant. The tasks may include e.g. performing a gate between remote qubits \cite{Baeumer2024a,Plathanam2025}, performing multiple gates between different pairs of qubits \cite{Beverland2022,Dur2025}, running a specific sub-circuit on a subset of qubits, performing an error syndrome read-out or an error correction step when dealing with encoded quantum information, performing logical gates on logical systems, use graph states to reduce the circuit depth \cite{Kaldenbach_2025} etc..
In the following we will consider the realization of multiple two-qubit gates between different pairs of qubits, and performing arbitrary permutations on a subset of qubits, as these are central elements repeatedly used within many quantum algorithms. Our methods are however not limited to these specific tasks, and one may find optimized entangled resource states that can assist to perform other tasks more efficiently, and in this sense also increase the power of a quantum processor. To this aim, we compare the required number of steps to perform a given task using the quantum processor without auxiliary entanglement, with the cost to perform this task when a certain auxiliary multipartite entangled state is available. If fewer steps are required, we say that the power of the quantum processor is increased. The overhead associated with a restricted interaction geometry for a conventional, gate-based quantum processing without auxiliary entanglement was analyzed in detail e.g. in Refs. \cite{cowtan2019,childs2019,Bapat2021,Bapat2023,Yuan_2025}.

Ref. \cite{Plathanam2025} analogously analyses the possible advantage of using dedicated auxiliary systems to perform long-ranged gates in a teleportation-based way, as compared to a fully gate-based approach, for different algorithms. However in contrast to our approach, Bell pairs are generated on demand in their setting, and no multipartite entanglement is utilized. Refs. \cite{Eldredge2020,Beverland2022,Devulapalli2024, Devulapalli2025a} also consider the potential gain of using teleportation-based gates for routing or performing permutations in settings with a restricted interaction geometry. They demonstrate the advantage over gate-based methods without auxiliary systems. Ref \cite{yang2023} discusses a dedicated entanglement bus, and the utilization of GHZ-states. Again, their settings are concerned with on-demand preparation of entangled auxiliary states.
Here we concentrate on the additional power enabled by some auxiliary, pre-prepared multipartite quantum state. We separate the preparation cost of these states from the actual cost of the quantum computation or specific task, which is possible since we use universal resource states that are independent from the desired algorithm to be implemented, and they can be generated before.

\subsection{Multiple two-qubit gates}
As described in Sec. \ref{Sec_Bellstateextraction}, one can generate $m_d={\cal O}(N^{(d-1)/d})$ Bell pairs in a single step from a dD-CS, and hence $m_d$ two-qubit gates between disjoint pairs of qubits can be realized in parallel in a single step using the scheme described in Sec. \ref{Telgate}. We now compare this to the cost to achieve the same task in different settings without utilizing pre-prepared auxiliary entanglement.

\subsubsection{Single long-ranged gate in a 1D and 2D geometry}
In the 1D interaction geometry with $N$ data- and $N$ auxiliary qubits (see Fig. \ref{Fig_1Dto2D}a), performing a single two-qubit gate between two (randomly chosen) qubits solely in a gate-based fashion requires ${\cal O}(N)$ steps: the average distance between two qubits is $N/2$, and in order to perform a two-qubit gate one needs to bring qubits next to each other with a series of on average $N/2$ SWAP operations, perform the desired gate, and SWAP the qubit back to its initial position. This procedure might be slightly improved by transporting both of the qubits and performing the corresponding SWAPS simultaneously, however still ${\cal O}(N)$ steps are required on average. The required number of steps is different in other underlying interaction geometries. For instance, a gate only requires ${\cal O}/\sqrt{N})$ steps in a 2D interaction geometry, as fewer steps are required to transport qubits next to each other.

This cost directly translates into the advantage of a dD-CS, where the advantage is the same for all $d\geq 2$.
Notice that for a single long-ranged gate there is essentially no advantage of our approach over on-demand implementation of teleportation-based gates \cite{Plathanam2025}. The required Bell pair can be generated in two additional steps, by preparing a 1D-CS by performing all the commuting CZ gates in one step, and measuring the intermediate qubits in $Y$-basis in a second step.

\subsubsection{Multiple long-ranged gates in a 1D geometry}
The cost to perform $m$ gates on disjoint pairs of qubits in a gate-based scenario is the same as for a single gate, since any permutation of $N$ qubits can be done in $N$ steps: all the SWAPs can be done simultaneously (in round $l$ between qubits $(k,k+1)$ for all $k$ odd [even] if $l$ is odd [even]) (see also \cite{Devulapalli2024}). As long as $m \leq m_d$, a dD-CS offers a constant advantage of ${\cal O}(N)$ over a gate-based approach. Since $m_d$ increases with $d$ (see Eq. (\ref{m_d})), a dD-CS with larger $d$ is more powerful to support the computation.

In a scenario where a quantum circuit consists of $m$ rounds, and in each round not only a single long-ranged teleportation-based gate is applied, but multiple short-ranged gates between data qubits are implemented directly in a gate-based way (and hence require to keep the locality structure to avoid overheads), the cost is given by $m$ times the cost to perform a single long-ranged gate, i.e. ${\cal O}(mN)$ (see also \cite{Plathanam2025}). Hence the advantage of using a dD-CS is given by ${\cal O}(N^{(2d-1)/d})$ as long as $m \leq m_d$.

\subsubsection{Multiple long-ranged gates in a bus-based setting}\label{bussetting}
Here we consider a gate-based set-up with all-to-all connectivity, but where only a single two-qubit gate can be performed per time step. This is the case when two-qubit gates use a joint bus, e.g. the motional mode in an ion-trap setup \cite{fazio2025} where Cirac-Zoller type gates are used \cite{Cirac&Zoller1995}. Hence two-qubit gates can only be performed sequentially, and it follows that the cost of a specific task is simply given by the number of involved two-qubit gates between arbitrary pairs of qubits.

Since there is no notion of distance in such a setting, also long-ranged gates have cost one. Performing $m$ two-qubit gates thus has cost $m$. It follows that a dD-CS has an advantage of $m$ if $m \leq m_d$, Eq. (\ref{m_d}). Again, dD-CSs are more powerful for larger $d$.
Notice that in such a bus-based setting, the advantage of our approach over on-demand teleportation-based implementations prevail, since the generation of the required 1D-CS involves ${\cal O}(N)$ gates and hence ${\cal O}(N)$ steps.

\subsection{Permutations on subsets of qubits}
As demonstrated in Appendix A, one can perform an arbitrary permutation on a subset of $m<m_d$ qubits in a system of size $N$ with constant overhead, using two copies of a dD-CS. In contrast, in a 1D geometry a gate-based approach without auxiliary entanglement requires ${\cal O}(N)$ steps, and a teleportation assisted approach ${\cal O}(m)$ steps \cite{Devulapalli2024}. Also in a bus-based setting (see Sec.\ref{bussetting}), one needs ${\cal O}(m)$ steps.

\section{Generation of $d$-dimensional cluster states in a 1D geometry}\label{Sec_Stategeneration1D}
We now turn to the generation of $d$-dimensional cluster state from an underlying 1D interaction geometry, i.e. when two-qubit gates between nearest neighbor auxiliary qubits are available. We start by describing a gate-based method to generate arbitrary graph states with circuit complexity of ${\cal O}(N^2)$. Essentially, SWAP gates between neighboring quibts can be used to move qubits around, while CZ gates allow one to add (or remove) edges after each such permutation. We SWAP qubit 1 to the last position, and apply a CZ gates to the qubit to the left at step $k$ if there is an edge $(1,k)$ in the graph. This involves $(N-1)$ SWAP gates and up to $(N-1)$ CZ gates, i.e. at most $(N-2)$ steps. We continue with qubit 2, which we swap only to the second last position, and perform intermediate CZ gates to add all edges $(2,k)$. This involves at most $2(N-2)$ steps. We continue in this way with qubits $3,4, \ldots N-1$, and after at most $2\sum_{k=1}^{N-1}(N-k)=N(N-1)$ steps the procedure finishes and an arbitrary graph state with all desired edges is generated.

In remainder of this section, we consider a measurement-assisted approach, where the generation complexity, i.e. the required number of steps, can be significantly reduced for certain target graph states if additional auxiliary qubits are available (see Fig. \ref{Fig_1Dto2D}). In the depicted set-up of Fig. \ref{Fig_1Dto2D}a, we actually consider in addition to the data qubits (green) two types of auxiliary systems: entangling systems that interact (set $E$, red) which are used to generate entanglement, and memory systems (set $M$, blue) where multipartite entangled states are stored and manipulated to perform gates between attached data qubits. We remark, however that such a distinction between entangling and memory systems is not necessary, and in fact a minimal setup would require only one auxiliary qubit associated with a data qubit. The additional memory systems are used to enhance preparation of multipartite entangled resource states, as we now demonstrate.

In this case, the elementary state that can be naturally generated between entangling qubits (system $E$) in one step is a 1D cluster state - or a  state where some of the edges are left out. This state can then further be modified via single-qubit measurements, and then used to alter the entanglement stored in the memory system $M$. In particular, one can generate a Bell state shared between any pair of entangling qubits from a 1D-CS by means of Pauli $Y$ and $Z$ measurements as described above, which can in turn be used to apply a CZ gate and hence add the corresponding edge to the graph state stored in $M$. By using at most $N(N-1)/2$ steps of this kind, one can generate arbitrary graph states in $M$ by adding all required edges. In a similar fashion, one can prepare dD-CSs in at most $dN$ steps. This is the number of edges in a regular $d$-dimensional rectangular lattice with periodic boundary conditions.

In the following, we discuss even more efficient schemes to generate dD-CSs by adding multiple edges in each step.
Fig. \ref{Fig_1Dto2D} illustrates the generation of a 2D-CS in an underlying 1D interaction geometry in $\sqrt{N} + 1$ steps. Similarly, Fig. \ref{Fig_1Dto3D} summarizes how to generate a 3D-CS in a 1D geometry in ${\cal O}(N^{2/3})$ steps. Further details are provided in Appendix B, where also the generation of dD-CSs in ${\cal O}(N^{(d-1)/d})$ steps is discussed.

\subsection{Gradually increasing processor power}
As we have shown above, the measurement-assisted generation of dD-CSs requires a number of steps that increases with $d$. If the available (or expected) idle time is known, one can fix the desired target state, e.g. a dD-CS with a large $d$, and prepare this state. A more interesting and relevant scenario is however given when the idle time, and hence the available time to generate supporting resource states, is unknown. Depending on the available resources, most importantly the available memory, different strategies are conceivable.

\subsubsection{Generation of additional resource states}
The simplest and most transparent setting is when multiple additional memory qubits (set $M$) are available. This allows one to simply store the already generated states, and start with the generation of an additional resource state. The strategy thereby depends on the expected available time, and the required support task. One can simply aim to generate certain Bell pairs, multiple 1D or 2D cluster states, or dD-CSs with higher $d$. The power of the available resources simply adds up, as all stored entangled states can be utilized to support the quantum computation. In this sense one can fill the whole available memory with resource states, that can be used to flexibly support a later computation. Without memory limitations, generating and storing 1D-CSs is already a good strategy, as each of these states can be generated in a single step, and be utilized to perform at least one long-distance teleportation-based two-qubit gate. Another natural strategy is to generate and store Bell states between different pairs of systems, however this requires a large memory to per system (up to $N-1$).

In case the memory is limited, one should adapt the strategy and work towards generating and storing multipartite entangled states. The main advantage of multipartite entanglement is the reduced memory requirement, and both the computational advantage as well as the required time to generate a dD-CS increases with $d$. However, generating a dD-CS requires a certain, fixed number of steps, and it is not straightforward to utilize some intermediate state where some edges are still missing. E.g. the zipper scheme of Ref. \cite{Freund2024} requires 2D-CSs, and does not work if edges are missing, or additional edges are present. In turn, we expect that the nested quantum switch construction put forward in \cite{Ramiro2026} profits from any available Bell pair, or any edge in the graph-state variant, and the corresponding computational power can hence be gradually increased directly.

When using dD-CSs as resources, we suggest to start by generating 1D-CSs in all available memories. If more idle time is available, one can work towards generating a 2D-CS in one set of memory qubits, and continue by upgrading one copy after another to a 2D-CS. One can continue in this way, and increase the dimension $d$ in one stored copy after another, generating 3D-CSs, 4D-CSs and so on.

\subsubsection{Modification of 2D-CS to generate 3D-CS}
The direct generation of a dD-CS as described above assumes that the initial state in $M$ is a product state, and one needs to generate all required entanglement, i.e. all edges. Here we discuss the case where one has generated some $d-1$ dimensional cluster state of $N$ qubits in $M$, and wants to modify or utilize it to generate a dD-CS state corresponding to dimension $d$. This is relevant if there is an unknown idle processor time available, and one wants to further increase the computational power of the auxiliary state - i.e. first generate a 1D-CS, then a 2D-CS, a 3D-CS etc..

A detailed description of how to generate a 3D-CS from an already established 2D-CS is given in Appendix C, which has the same scaling of ${\cal O}(N^{2/3})$ as the direct generation of a 3D-CS from a product state.

We finally remark that other underlying interaction geometries lead to a reduced cost to generate dD-CSs. In appendix D we show how to generate dD-CSs in a 2D geometry in ${\cal O}(N^{(d-2)/d})$ steps.

\section{Summary and outlook}\label{Sec_Summary}
We have shown that quantum processor architecture that includes both data and auxiliary qubits allows one not only to overcome limitation in intrinsic connectivity due to geometrical constraints \cite{Dur2025}, e.g. nearest-neighbor couplings in a 1D or 2D geometry, but also to store computational power in auxiliary entangled states. We have illustrated this by considering cluster states corresponding to rectangular lattices in $d$-dimensions, where we showed that the number of long-ranged gates that can be flexibly implemented by consuming the auxiliary resource state increases with the dimension $d$, and is given by ${\cal O}(N^{(d-1)/d}$. In turn, the generation of the resource states requires a number of steps that increases with $d$ in a set-up with limited connectivity. For example, in an underlying 1D nearest neighbor interaction geometry, ${\cal O}(N^{(d-1)/d})$ steps are required to generate such a $d$-dimensional cluster state. Given that resource states can be generated beforehand, even before the computation starts or even before it is known which computation should be performed, one can use processor idle times to generate such resource states that can later be used to enhance the power of the quantum processor. What is more, the power can be gradually increased if more idle time is available. We have illustrated this by showing how to generate $d+1$-dimensional resource cluster states from $d$-dimensional ones using the underlying 1D interaction geometry.

If more memory qubits in set $M$ are available, multiple copies of resource states or different families of resource states can be generated and stored during idle times. These states can then be used to assist and speed up computations at a later stage. More auxiliary qubits allow one to store more resource states, which in turn alow one then to perform a larger number of gates or more complex auxiliary circuits in short time (often within a single step) at a later stage. The usage of multipartite entangled states is beneficial if memory is limited, and flexibility is an issue. The generation of Bell states from a multi-partite entangled resource states requires some overhead, in terms of auxiliary systems that need to be measured. The direct usage of a Bell state shared between modules to perform a two-qubit gate between the corresponding data qubits, in turn, has no overhead. However, if one wants to guarantee that two-qubit gates between arbitrary pairs of systems can be performed, one needs to store all $N(N-1)/2$ different types of Bell states, which requires $(N-1)$ qubits per node. As illustrated with examples of $d$-dimensional cluster states, with only a single qubit stored qubit per data qubit, multiple two-qubit gates can be realized, and flexible chosen.

Furthermore, the strict separation between performing single-qubit gates on data qubits directly, and all two-qubit and multi-qubit gates via teleportation-based gates or measurement-assisted methods is not necessary. One can also perform some of the two-qubit gates directly on demand, or specifically generate the required entanglement. Stored multi-partite entanglement may be utilized only if a direct implementation is not possible, or too costly. In this sense our approach can also be used to assist an otherwise standard, gate-based quantum computer. In principle, also a flexible assignment of qubits to data- and auxiliary qubits is possible.
Clearly, the method is not limited to qubits, but can also be employed when qudits are used to process or store quantum information. Multipartite entangled qudit states can even offer additional flexibility, and allow for more efficient generation of desired target states to perform different tasks.

We envision that a quantum computation then happens in such a way that entanglement stored in different resource states is flexibly used to assist computation on data qubits, and whenever parts of the quantum processor are not actively needed in the ongoing computation, entanglement is generated to either refresh partly consumed resource states, enhance their power, or generate new ones. In this sense, available quantum memory is used not only to store quantum data, but can actively assist quantum computations and speed them up. Such an active and supporting role of the available memory is in stark contrast to classical computation, and promises new possibilities in quantum processor and quantum computer design.

\subsection{Alternative resource states}
We note that other multipartite state structures beyond $d$-dimensional cluster states are also suitable as resource states to improve the computational power of quantum processors. As discussed in \cite{Ramiro2026}, a multipartite entangled graph state corresponding to a hypercube of dimension $d=\log_2N$ can act as an (almost) universal quantum switch, i.e. allows one to obtain ${\cal O}(N/(\log N)^2)$ parallel Bell states in an (almost) arbitrary configuration using only one qubit per module. A hypercube of dimension $d$ is defined as a graph with $2^d$ vertices, where two vertices (labeled in binary notation) are connected by an edge if their corresponding bit strings have Hamming distance one, i.e. differ by exactly one bit. The local vertex degree is $d$ and the average path length between two vertices is $d/2$. A Bell state between two nodes can be generated by measuring all qubits along a connecting path (of length $d/2$) in $Y$, and measuring all $d$ neighbors of each of the vertices along the path in $Z$.
Similarly, by using the same geometry but separate Bell pairs for each edge rather than a graph state, one obtains a set-up with $d=\log N$ auxiliary qubits per module, but where any configuration of $N/2$ disjoint Bell pairs can be generated.
Such a state can be obtained from a 1D interaction geometry in $N$ steps, and one can gradually increase its power by adding more and more edges. This is particularly useful in a dynamical setting, where idle times during computations are used to enhance resource states. We also remark that auxiliary resource states that are not fully consumed can be patched, and restored to its initial functionality. This can be done by adding the required edges of all qubits that have been measured during the generation process of long-distance Bell states or other resource states. The exact number of required steps depends on the measured qubits. A relevant question in this context is to identify for a given number of auxiliary qubits sets of resource states that can be flexible used, and efficiently generated in a given geometry.

While we have concentrated on flexibly generating multiple Bell states from a given multipartite entangled resource states to enable multiple long-ranged gates and overcome limitations due to localized interaction geometry, a much more broader assistance of quantum computations enabled by stored auxiliary entanglement is conceivable. As outlined in \cite{rierasabat2024}, one can also perform multi-qubit gates or even whole Clifford circuits on subsets of qubits, using GHZ states or specific graph- or stabilizer states. It is also conceivable to perform stabilizer measurements on subsystems, or error correction steps utilizing entanglement, and generate the required task-specific resource states from some universal resource such as a dD-CS. We remark that in \cite{Freund2025} it was shown how to directly generate graph states from 2D-cluster states, which in turn can serve as a resource for different tasks. Recall that any Clifford circuit acting on $m$ qubits is associated with a graph state of size $2m$, and can be implemented using such a graph state in a single step. Our general approach is however not limited to these specific tasks or resources, and it would be interesting to identify other processes where entanglement can be beneficial.

\subsection{Noise and imperfections}
Another relevant aspect is concerned with noise and imperfections. On the one hand, noisy multipartite resource states lead to imperfect gates, where the achievable fidelity depends on the traversed distance. As shown in \cite{MorRuiz2024}, the quality of generated Bell states depends on the dimension of the underlying dD-CS, and for a fixed total system size $N$ there is some optimal dimension $d$, where the optimal $d$ increases with $N$. In \cite{Dur2025} it was shown how this translates into achievable gate fidelities.
One can further enhance teleportation-based gate fidelities in a 2D-CS by using encodings \cite{Romanova2026}. This allows one to deal with noise from state generation, but also from imperfect memory when storing pre-prepared entangled states. In practice, these stored resource states or parts of them will have to be refreshed. Obtaining proper strategies for this in realistic settings is an interesting future task.

On the other hand, our approach is also applicable in fault-tolerant quantum computation set-ups. Remote gates between encoded (logical) quantum systems can be implemented using encoded entangled states. The proper choice of error correction codes and quantum information processing allows one to utilize auxiliary entanglement also in such settings, e.g. to perform long-rang logical gates between distant logical systems. In principle, auxiliary entanglement can also be used to directly perform tasks such as direct, measurement-assisted syndrome readout or error correction, or to use Magic states to remotely perform non-Clifford operations within our setting.

Our approach shows that quantum memory can be a valuable resource in quantum computation beyond simply storing quantum data, and can in fact take an active role to increase the computational power of a quantum computer. By continuously generating, maintaining or upgrading auxiliary entangled states using parts of the quantum processor that are otherwise unused, these states can be utilized to enhance the computational power of quantum processor, and reduce overheads for different tasks.

\section*{Acknowledgments}
This research was funded in whole or in part by the Austrian Science Fund (FWF) 10.55776/P36009, 10.55776/P36010, 10.55776/PAT1710825 and 10.55776/COE1. For open access purposes, the author has applied a CC BY public copyright license to any author accepted manuscript version arising from this submission. Finanziert von der Europ\"aischen Union - NextGenerationEU.
The method described here was filed as a patent.

\begin{widetext}

\begin{figure}[ht]
    \centering
    \includegraphics[width=\columnwidth]{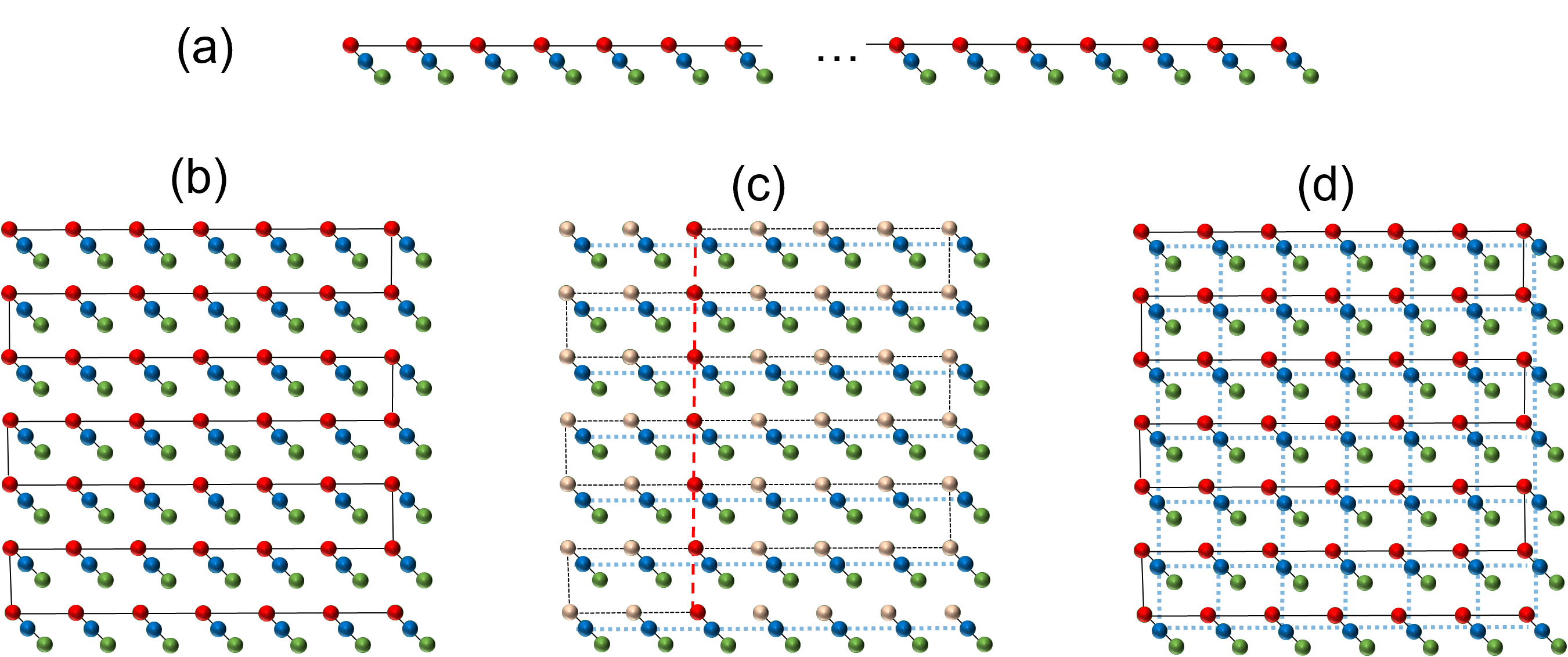}
    \caption{\label{Fig_1Dto2D} (a) 1D interaction geometry, where to each data qubit (green) one memory qubit (blue) is attached, which in turn is coupled to an entangling qubit (red). The entangling qubits interact according to a 1D nearest-neighbor interaction geometry. Black lines indicate the interaction pattern, i.e. the availability of two-qubit CZ-gates to generate entanglement. (b) Qubits can be (virtually) arranged in a snake-like structure on a $\sqrt N \times \sqrt N$ grid. (c) Entanglement is be generated according to the interaction geometry among all horizontal lines, and transferred to the auxilary system (blue qubits). One can also generate a 1D cluster state and transform it via $Y$-measurements (light red qubits) to form a 1D cluster state on one vertical line of qubits (red). (d) The entanglement of the vertical lines can be transferred/added to the entanglement of the auxiliary system, thereby generating a 2D cluster state on the blue auxiliary qubits after $\sqrt N+1$ rounds.
   }
\end{figure}

\begin{figure}[ht]
    \centering
    \includegraphics[width=\columnwidth]{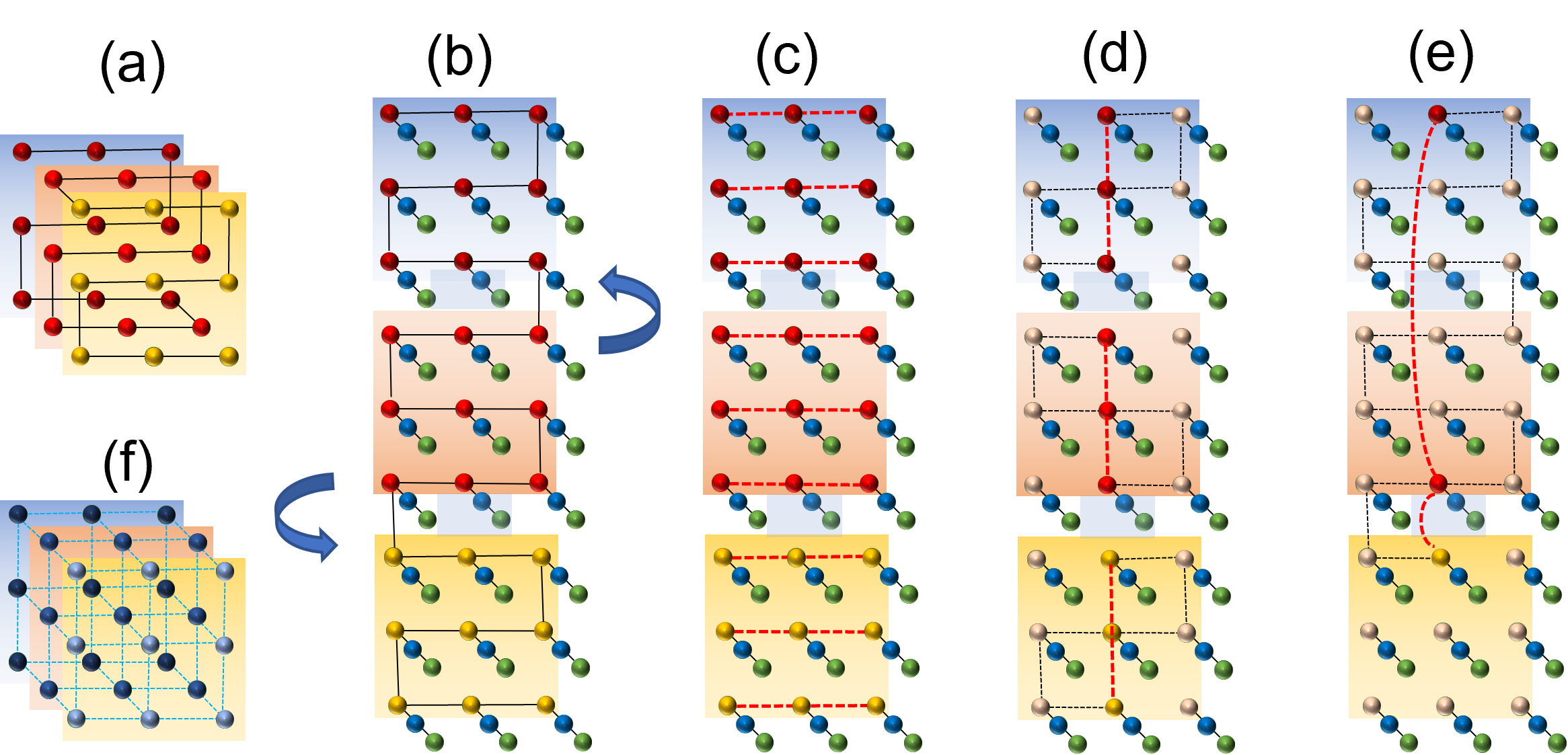}
    \caption{\label{Fig_1Dto3D} (a) Entangling qubits $E$ are (virtually) arranged on a 3D lattice of size $3 \times 3 \times 3$, which is obtained from a rectangular (virtual) 2D structure (b) on a rectangular lattice of size $3 \times 9$ by folding. Notice the snake-like arrangement of the underlying 1D interaction geometry (black lines). (c) The available gates in $E$ can be used to generate all required edges in $x$ direction in one step. (d) All edges in $y$ direction in all planes, corresponding to one fixed $x$ coordinate ($x=2$ here), can be generated in one step by performing the indicated $CZ$ gates, and $Y$ measurements among the pathes. To generate all required edges in $y$-direction, one needs hence three steps. (e) Edges in $z$ direction for a fixed $x,y$ coordinate ($x=2,y=1$ here) can be generated in one step in a similar way. In order to generate all required $z$-edges, one needs 9 steps. (f) Resulting 3D cluster state stored in $M$, where for clarity (a) only shows the entangling system $E$ and (f) only the memory system $M$.
   }
\end{figure}

\end{widetext}

\bibliographystyle{apsrev4-2}
\bibliography{ResourceStatePumping.bib}

\newpage
\section*{}
\newpage

\section*{Appendix A: Permutations on subsets of qubits}
Performing arbitrary permutations on subsets of qubits is an important element to overcome geometric interaction restrictions. Here we discuss the complexity of this task with and without auxiliary entangled states.

We first show that one can perform arbitrary permutations on subsets of qubits with constant overhead, utilizing multipartite entangled states as resource. In particular, two copies of a $N$ qubit dD-CS allow one to perform an arbitrary permutation on $m_d={\cal O}(N^{(d-1)/d})$ qubits (set $B$) in three steps, by generating Bell states between the source position $k_S$ for each qubit in set $B$, and its target position $k_T$ which can be used to teleport each of the corresponding data qubits to the appropriate position. This can be done by generating Bell states between the source position qubits $k^{(1)}_S$ to auxiliary qubits $k^{(1)}_{C}$ in another set $C$ distinct from $B$ from the first copy of the dD-CS, and Bell state between $j^{(2)}_k$ and the target position $k^{(2)}_T$ from the second copy (step 1). The Bell states $(k^{(1)}_S, k^{(2)}_T)$ can then be generated by performing CZs a Bell measurement on qubits $k^{(1)}_{C},k^{(2)}_{C}$ (or equivalently a CZ followed by $Y$-measurements on both qubits). At the same time, a Bell state measurement between $k^{(1)}_S$ and the corresponding data qubit is performed (step 2). In step 3, $k^{(2)}_T$ is SWAPed with the corresponding data qubit, leaving all data qubits in set $B$ at the permuted positions.

In a 1D interaction geometry, a gate-based implementation of an arbitrary permutation on a subset of $m$ qubits on a system of size $N$ requires ${\cal O}(N)$ steps \cite{Devulapalli2024}. Even when assisted by teleportation, certain permutations still require ${\cal O}(m)$ steps, e.g. permuting $(k,N-k)$ for $k=1,\ldots,m/2$ \cite{Devulapalli2024}.

In a bus-based setting where gates can only be done sequentially, see Sec.\ref{bussetting}, performing an arbitrary permutation on a subset of $m$ qubits has a worst-case cost of $m-1$. This follows from the fact that one can sort a list by simply swapping the first element to the correct position, then the second element and so on, where no swapping is required for the last element.

\section*{Appendix B: Generating cluster states in 1D geometry}
\subsection*{2D cluster state in 1D geometry}\label{Sec1Dto2D}
We consider the generation of a 2D-CS in an underlying 1D interaction geometry, depictured in Fig. \ref{Fig_1Dto2D}. We virtually arrange the $N$ modules in a snake-like structure, forming a 2D grid of size $\sqrt{N} \times \sqrt{N}$ (see Fig. \ref{Fig_1Dto2D}b), where each line indicates the possibility to apply an entangling gate. The qubits in $M$ and $E$ are all initialized in state $|+\rangle$. (i) In a first step, we apply all gates that correspond to horizontal lines in our virtual arrangement on qubits in $E$, but not the ones corresponding to vertical lines. This produces $\sqrt N$ copies of 1D cluster states of size $\sqrt{N}$ in $E$, which can then be used to apply multiple CZ gates on the memory qubits in $M$, thereby generating all horizontal edges of the target 2D-CS. (ii) In the next step, a 1D cluster state generated in $E$ and modified by means of $Z$ and $Y$ measurements to form a 1D cluster state on a subset $E_k$ of qubits, arranged on the $k^{\rm th}$ vertical line. This is done by leaving all qubits in set $E_k$ untouched, while measuring all qubits on the path between the first and the last qubit in $E_k$ in $Y$. The remaining qubits at the beginning an end of the chain are measured in $Z$ (or alternatively, a shorter 1D cluster state is generated on the entangling qubits to start with). The resulting vertical 1D cluster state can then be used to generate all vertical edges in column $k$ on the auxiliary graph state stored in $M$ in one step (see Fig. \ref{Fig_1Dto2D}c). By repeating this step for all columns, i.e. $\sqrt{N}$ times, one generates a 2D-CS among the memory qubits (see Fig. \ref{Fig_1Dto2D}d). This involves $\sqrt{N} + 1$ steps.

\subsection*{$d$-dimensional cluster state in 1D geometry}\label{Sec1Dto3D}
The generation of 3D-CS (or higher dimensional one) takes place in a similar fashion. We virtually arrange the $N$ modules in a cube of size $N^{1/3} \times N^{1/3} \times N^{1/3}$. We use a snake-like structure, this time of size $N^{1/3} \times N^{2/3}$, see Fig. \ref{Fig_1Dto3D}. The first square of size $N^{1/3} \times N^{1/3}$ forms the first $xy$ plane of the cube, the second square then the second $xy$ plane with $z=2$ and so on, where again the planes are arranged in a snake-like way, see Figs. \ref{Fig_1Dto3D}ab. Notice that there is always only a single connection between two neighboring planes, and the position alters between top and bottom. (i) By applying all gates corresponding to edges in $x$-direction in $E$, but not the ones in $y$ and $z$ direction, one can use the resulting state in $E$ to generate all edges in $x$-direction of the target 3D-CS in $M$ in a single step, as illustrated in figure \ref{Fig_1Dto3D}c. (ii) One can generate in each plane the $k^{\rm th}$ line in $y$ direction (i.e. $y$ coordinate is fixed to $k$) in the same way as described in Sec. \ref{Sec1Dto2D}, step (ii) - see \ref{Fig_1Dto3D}d. Notice that each of the planes can be accessed independently (no connecting gates between the planes are applied), and hence the edges of all $N^{1/3}$ vertical lines with $y$ coordinate $k$ can be generated simultaneously. By repeating this step for all $k$, i.e. $N^{1/3}$ times, one can generate all edges in $y$-direction of the target 3D-CS in $M$. (iii) Edges in $z$ direction can be generated in a similar way, however due to the chosen arrangement there is only a single connection between two planes. Hence only a singe line in $z$-direction (corresponding to a 1D cluster state between all qubits with fixed $x,y$ coordinates, and arbitrary $z$ coordinate), can be generated from an initial 1D cluster state in $E$, see figure \ref{Fig_1Dto3D}e. One needs to repeat this step $N^{2/3}$ times to obtain all required lines, i.e. all edges, in $z$ direction of the 3D-CS in $M$. The required number of steps to prepare a 3D-CS is hence $N^{2/3} + N^{1/3} + 1 = {\cal O}(N^{2/3})$.

One can prepare dD-CSs in a similar fashion, arranging the initial 1D cluster state on a $d$-dimensional cube. In 4D, this would correspond to virtually arranging the chain on a $N^{1/4} \times N^{3/4}$ grid, then folding this in a similar way as described above to have $N^{1/2}$ 3D cubes which can then be folded (in a snake-like fashion) to form a 4D cube of size $N^{1/4}$ in each dimension. For the first three dimensions, parallel edges can be generated: all edges in first dimension in a single step, all edges in the second dimension in $N^{1/4}$ steps, the edges in the third dimension in $N^{2/4}$ steps. In the fourth dimension, only a single line can be generated from one 1D cluster state, requiring then in total $N^{3/4}$ steps. A similar argument shows that to generate a cluster state of dimension $d$, one needs $\sum_{k=1}^{d-1} N^{k/d}+1 = {\cal O}(N^{(d-1)/d})$ steps.

\section*{Appendix C: Modification of 2D-CS to generate 3D-CS}
Here we demonstrate how to generate a 3D-CS from a 2D-CS. We first remark that the same procedure that allows one to use entanglement in $E$ to add an edge in $M$ actually also allows one to remove an edge that is already there, since CZ$^2$ is the identity. In the following we discuss the generation of a 3D cluster state from a 2D one to illustrate our procedure. The argument can be generalized to arbitrary $d$. We consider two change the virtual arrangement of qubits from 2D to 3D by grouping the $\sqrt N \times \sqrt N$ qubits in blocks of size $m=N^{1/3} \times N^{1/3}$, with $N^{1/6}$ of these blocks in $x$ and $y$ direction. Notice that the available physical connections remain unchanged, i.e. correspond to 1D snake-like structure within the virtual 2D setting, see Fig. \ref{Fig_1Dto2D}. If we imagine now again a snake-like pattern between the blocks, they can be folded to form a 3D cube of size $N^{1/3} \times N^{1/3} \times N^{1/3}$. Notice that in this way, many of the required edges of the 3D-CS are already there if we start with a 2D-CS , in particular all edges in the $xy$ planes of the 3D arrangement. However most of the edges in the third dimension are missing, and some of the edges which are already there need to be removed. The underlying 1D geometry determines the cost to add- or remove edges between different blocks, and also which of them can be generated simultaneously in one step. Horizontal edges can be generated in parallel and independently, while only one vertical line can be generated in one step. Removing edges in $y$ direction --to form the snake-like pattern of arranged blocks-- needs $N^{1/2}-N^{1/3}$ steps.  Adding the required edges in third dimension requires $2N^{2/3}$ steps for the vertical ones (of the initial 2D lattice) and $N^{1/3}$ for the horizontal ones. The factor of two stems from the fact that the left and right collumn of Blocks need to be treated separately and hence sequentially. While this strategy has the same scaling of ${\cal O}(N^{2/3})$ of required steps, it actually takes slightly longer than generating a 3D-CS from a product state (see Sec. \ref{Sec1Dto2D}).

Clearly, one can use the already established 2D-CS in a more clever way that actually reduces the number of required steps to generate a 3D-CS. One possibility is to modify the 2D-CS by means of measurements to establish multiple edges that are required for the 3D-CS. In particular, one can obtain $O(\sqrt N)$ edges between flexibly chosen pairs of qubits. From there, one uses the strategy outlined in Sec. \ref{Sec1Dto3D}, leaving out steps for edges that are already established using the initial 2D-CS. Notice, however, that the possible assistance due to the already generated 2D-CS is limited, and still $O(N^{2/3})$ additional steps are required to end up with a 3D-CS.
\newline

\section*{Appendix D: Generation of $d$-dimensional cluster states in a 2D geometry}\label{Sec_Stategeneration2D}
\begin{widetext}

\begin{figure}[ht]
    \centering
    \includegraphics[width=\columnwidth]{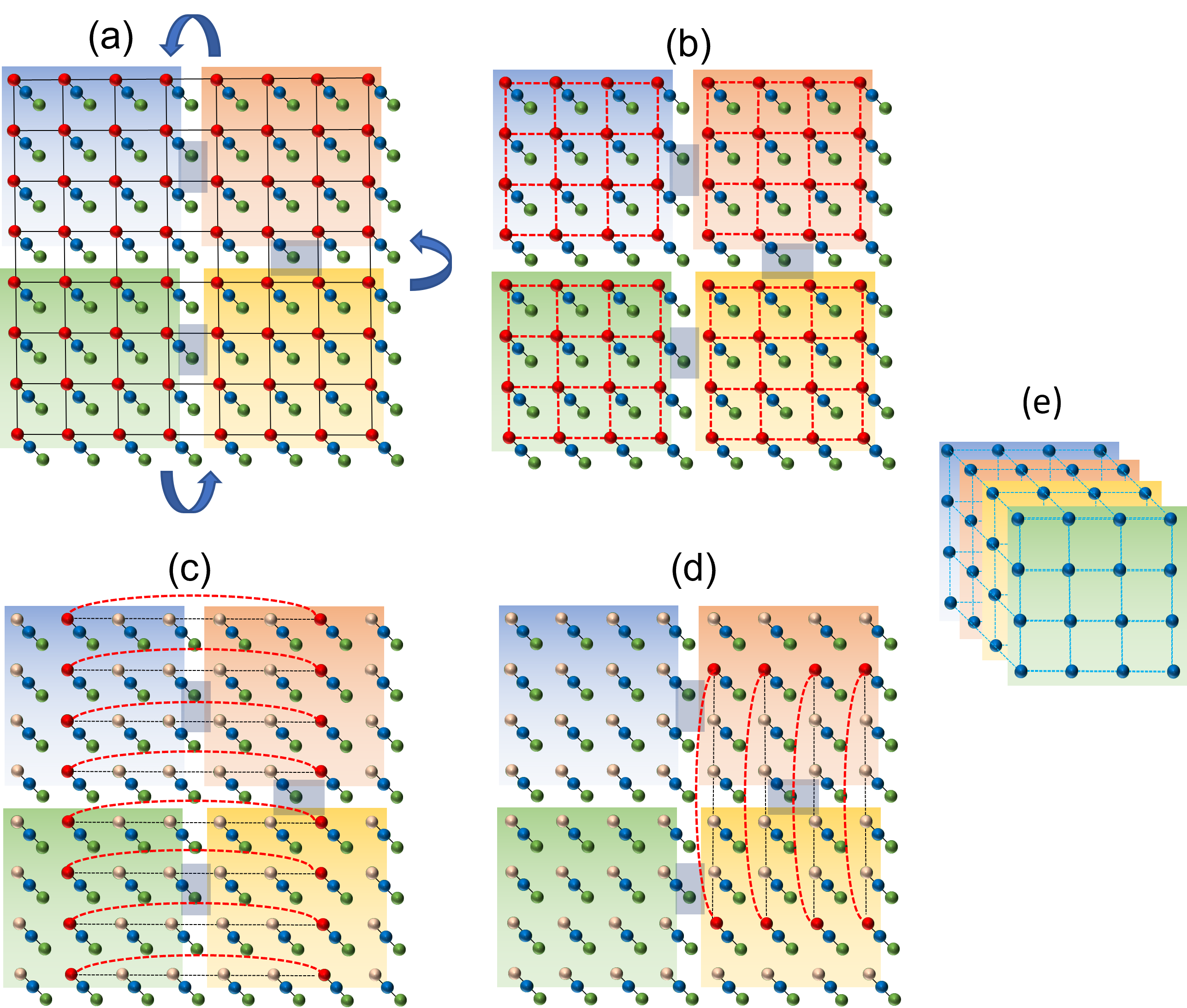}
    \caption{\label{Fig_2Dto3D} (a) System of size $8 \times 8$ with underlying 2D interaction geometry, where available interactions are indicated by black edges. The entangling qubits are virtually arranged into elementary squares of size $4 \times 4$ (blue/red/yellow/green), which are folded to form a 3D-CS (cube (see (e)). (b) All edges in the $x-y$ plane of the cube can be generated in a single step. (c) Edges in $z$-direction between neighboring planes with fixed $y_k$ that are connected horizontally in the underlying 2D geometry can be generated in a single step, by producing 1D cluster states and measuring intermediate qubits in the $Y$-basis. Figure illustrates the case $y=2$, and this step needs to be repeated for different $y_k \in\{1,2,3,4\}$. (d) Edges between neighboring qubits with fixed $x_k$ that are connected vertically in the underlying 2D geometry can be generated in parallel. Figure shows $x=2$, and this step needs to be repeated for $x_k\in\{1,2,3,4,\}$. Notice that the entanglement generated between the auxiliary qubits $E$ (red) is used to generate edges on the system $M$ (blue) to finally form a 3D-CS between, see figure (f).
   }
\end{figure}

\end{widetext}

One can also use other underlying interaction geometries to perform resource state pumping. Consider for example a set-up with a rectangular 2D interaction geometry of size $m \times m$. As we show below, a more efficient generation of a dD-CS is possible in this case. Similarly as in the 1D geometry case, we virtually arrange elementary modules in a snake-like fashion, where we consider $\sqrt n \times \sqrt n$ squares of size $n \times n$ as elementary building blocks. These $n$ squares are folded to obtain a cube of size $n^3$, which is illustrated in figure \ref{Fig_2Dto3D} for $m=8$ and $n=4$. Notice that this is guaranteed to work for system sizes of $N=k^6$ and integer $k>1$, where $m=k^3$ (side length of 2D geometry), $n=k^2$ (length of cube, and equivalently of the elementary square) and $\sqrt{n}=k$. For other values of $N$, rectangular geometries, possibly with some unused auxiliary modules, need to be considered. In the following, we use $x$,$y$ and $z$ coordinates to label the qubits in the 3D-CS. In the following we describe how to obtain all required edges for the target 3D-CS stored in the memory system $M$. Similarly as in Sec. \ref{Sec_Stategeneration1D}, we assume that the available nearest-neighbor gates are used to generate a graph state in system $E$, which is then manipulated by means of measurements to establish a certain sub-set of edges required in the 3D-CS. These edges are added to the memory system $M$ by using the entanglement in $E$ to perform teleportation-based CZ-gates as outlined in Sec. \ref{Sec_Background}.
All required connections of the 3d-CS (cube) within the $x-y$ plane can be obtained in a single step. The required $z$-connections between neighboring planes of the cube need to be done separately, where we distinguish between horizontal connections and vertical connections in the underlying 2D geometry due to the snake-like arrangement. For any column $y_k$, i.e. all qubits with fixed $y_k$ coordinate, all $z$-connections between neighboring elementary squares that correspond to horizontal connections in the underlying 2D geometry can be done in a single step by generating 1D-CS along horizontal lines, and measuring intermediate qubits in $Y$. Similarly, for any row with fixed $x_k$ coordinate, all $z$-connections to neighboring squares (the ones at the turning points of the snake-like arrangement) can be done in one step. Hence, a total of $2n = 2 N^{1/3}$ steps are required to obtain all $z$-connections, leading to $2 N^{1/3}+1 = {\cal O}(N^{1/3})$ steps in total. This should be compared to the ${\cal O}(N^{2/3})$ required steps for an underlying 1D geometry.

This method can be extended to generate a dD-CS in a 2D geometry, by virtually arranging squares of size $N^{1/d} \times N^{1/d}$ in a snake like fashion, and folding them appropriately. The cost is given by ${\cal O}(N^{(d-2)/d})$.
We conjecture that a dD-CS can be generated in an underlying $3D$ geometry with $d\geq \tilde d$ in ${\cal O}((d-3)/d)$ steps.

\newpage

\end{document}